\documentclass[a4paper,11pt]{article}
\usepackage{jcappub} 

\title{\boldmath Novel Simulation Framework for Analyzing Cosmic Ray Particle Distributions at a Global Scale}

\author[a,b]{O. Sarajlic,}
\author[a]{X. He}
\affiliation[a]{Department of Physics and Astronomy, Georgia State University,\\
Atlanta GA, United States}
\affiliation[b]{Continuing Education, Emory University,\\
Atlanta GA, Unites States}

\emailAdd{xhe@gsu.edu}

\abstract{
Cosmic ray measurements have inspired numerous interesting applications over several decades worldwide. These applications encompass non-invasive cosmic ray muon tomography, which enables the imaging of concealed dense objects or structures, the monitoring of area-averaged soil moisture with cosmic ray neutrons in agriculture and climate studies, real-time monitoring of the dynamical changes of the space and earth weather, etc. The demand for a quantitative characterization of cosmic ray shower particles near the Earth's surface is substantial, as it provides realistic particle spectra and rates for these diverse applications. In this study, we introduce Earth Cosmic Ray Shower (ECRS), a GEANT4-based software designed to simulate cosmic ray particle interactions in the atmosphere. ECRS incorporates the U.S. Standard Atmospheric Model and integrates a time-dependent geomagnetic field based on the Tsyganenko and IGRF models. Additionally, we present two case studies illustrating variations in the location-dependent average particle energy for muons, electrons, neutrons, and gammas at sea level. An outlook of this project is provided toward the conclusion.
}

\begin{document}
\maketitle
\flushbottom


\section{Introduction} \label{sec:intro}

Since the discovery of the cosmic ray radiation by Victor Hess in 1912, the study of the cosmic ray radiation has played a significant role in the advancement of subatomic particle physics and inspired the rapid development of particle accelerators for searching the fundamental building blocks of matter \cite{Martin:NP}. The measurement of ultra-high energy cosmic rays ($>10^{18}$ eV) \cite{PierreAuger} is still a worldwide effort to study the origin of these particles and the connection to the astrophysical events. 

Over the past few decades, many interesting applications of cosmic ray measurements have been explored, which include the cosmic ray muon tomography for homeland security \cite{Nature:Muons}, volcanic activity monitoring and nuclear reactor core monitoring \cite{PRL:Fukushima}, soil texture and moisture monitoring \cite{Dopper:soil_monitor}, cosmic-ray dose monitoring network \cite{Liu:rad,Liu:monitoring}, and the discovery of a big void of Khufu's pyramid \cite{Pyramid_muon,Muon_tomography}. At the same time, numerous studies have reported correlations between the dynamical changes in the earth atmosphere and cosmic ray flux variations measured at sea level~\cite{kirkby:climate,lu:correlation,Ollila2012ChangesIC,shaviv:climate}. 

Exposure to cosmic rays poses a health risk especially to astronauts whose missions are beyond the reach of Earth's magnetic field and atmosphere \cite{SpaceRadiation,schimmerling:radiation}. Aircrew and passengers are exposed to a factor of 30 on average more cosmic ionizing radiation on every flight in comparison to the level of cosmic ray radiation at the ground level.
Flying short flights at low latitudes are preferred due to less radiation exposure because of the greater amount of radiation shielding provided by the Earth’s magnetic field and the air mass. The effect of the magnetic field shielding is maximum near the equator while gradually decreasing to zero near the polar regions \cite{un:radiation}.
On average, the aircrew can be exposed to 0.2 mSv to 5 mSv per year
\cite{osullivan:radiation}. 
Also, the dose rate over the polar region is strongly influenced by the solar activity. A much greater rate is seen during solar minimum for the flights over polar region than during solar maximum with increase of $\sim$50$\%$ at 15,000 m flight altitude \cite{osullivan:radiation}. 
Given these conditions, the exposure of the cosmic ray radiation is regarded as an occupational hazard for the airline crew \cite{ICRP,RadDose}. 


At the surface of the earth, on average, more than $80$\% of secondary cosmic ray shower particles are muons with a small percentage of neutrons, electrons and gamma ray photons. A human body receives $\sim$0.4 mSv yearly from cosmic ray radiation at sea level. However, the energy of the secondary cosmic ray muons at the surface varies with geoposition because of the Earth's magnetic field. Very low energy muons ($< 50$ MeV) can enter human body and decay to energetic electrons which create a series of ionizations to body tissue. The details of this study has not been fully explored.
 



There have been several models such as CARI-7 that estimates the dose of space radiation \cite{CARI7}, CRAC::CRII model based on CORSIKA and FLUKA Monte Carlo tools that computes the ionization by cosmic rays in the atmosphere \cite{Usoskin2010,Usoskin2011}, GEANT4 CRY generator for simulating the muon interactions within the detector to form 3D tomographic images of high-Z materials \cite{MuonTomo}, and others. At present, the CORSIKA package \cite{heck:corsika} is the most widely used code for cosmic ray simulations. 
CORSIKA's primary particle energy ranges from low energy ($10^{6}$ eV) up to ultra-high energy cosmic rays ($10^{20}$ eV), which is developed mainly for reconstructing the full energy of the very high energetic primary particles and their directional origin.

All these software packages are designed for specific studies of cosmic rays but are lacking the flexibility, and for the most part, ability to explore the full cosmic ray showers in the entire atmosphere with varying atmospheric air mass distribution and the geomagnetic fields in order to meet the needs of growing applications of the cosmic ray measurements.

Given these important topics of cosmic ray study and the associated applications aforementioned, it is essential to explore and quantify cosmic ray shower characteristics in the Earth's atmosphere, at sea level, and underground. 
In this study, we introduce Earth Cosmic Ray Shower (ECRS), a GEANT4-based software designed to simulate cosmic ray particle interactions in the atmosphere~\cite{xhe_git}. ECRS incorporates the U.S. Standard Atmospheric Model and integrates a time-dependent geomagnetic field based on the Tsyganenko and IGRF models.
The first version of this simulation software was developed at Georgia State University in 2007 \cite{sanjeewa:ecrs}. A significant upgrades of this software has been achieved over the past few years in several aspects, which include the improved and accurate atmospheric air density profile, expanded range of the geomagnetic field implementation, and the tracking of the creation of the secondary cosmic ray shower particles in the atmosphere, at sea level and underground. 

ECRS simulation provides a detailed and realistic representation of how secondary cosmic ray particles interact with the Earth's atmosphere, taking into account factors like altitude, latitude, longitude, and magnetic field strength. By varying the geomagnetic field following the IGRF model \cite{finlay:igrf}, one could also explore the solar modulation of the cosmic ray showers quantitatively with the simulation introduced in this work. This accuracy is crucial for understanding the distribution of cosmic ray induced radiation at different locations on Earth.
The flexibility of configuring ECRS simulation with the time-dependent geomagnetic field enables it to incorporate solar cycle effects and meteorologicaly induced fluctuations in cosmic ray flux. This is significant because cosmic ray intensity varies with solar activity, and meteorological conditions can influence the flux reaching the Earth's surface. 
Additionally, ECRS can serve as a benchmark for validating analytical models used in environmental monitoring studies. While other Monte Carlo simulations such as the ones used for environmental radiation monitoring \cite{Liu:dose_cal} or assessment of atmospheric profile at mountain altitudes~\cite{Brall:mountain_alt, Brall:moisture_snow} that provide simplifications for practical calculations, may not capture the full complexity of cosmic ray interactions unlike ECRS simulation. 

This paper is organized as following: in Section \ref{sec:simulation}, we describe the details of the ECRS simulation framework which is followed with two case studies, described in Section \ref{sec:result}, of the preliminary results of the cosmic ray particle energy distributions for muons, neutron, electrons and gamma at a global scale.  A summary and outlook is given in Section \ref{sec:summary}.

\section{Cosmic Ray Shower Simulation}
\label{sec:simulation}

The Earth Cosmic Ray Shower (ECRS) simulation is based on the GEANT4 software toolkit developed at CERN \cite{agostinneli:simulation,allison:geant4}. 
Specifically designed to systematically model cosmic ray showers in the Earth's atmosphere, ECRS incorporates the effects of variations in the geomagnetic field.  
GEANT4, GEometry ANd Tracking, is a toolkit used for simulating the passage of particles through matter using Monte Carlo methods, which has been widely tested and applied in high-energy nuclear and particle physics, space radiation, and medical radiation sciences for more than three decades.

The Earth's atmosphere functions as a vast particle detector, where cosmic ray particles interact with air molecules, generating showers of secondary particles that cascade towards the Earth's surface. The atmospheric air density profile is modeled following the U. S. Standard Atmospheric Model \cite{nasaDenMod}. To capture all of the cosmic ray shower activities in the atmosphere, we limit the depth of atmosphere to 100 km which is divided evenly into 100 layers. 
The troposphere extends from the Earth's surface to 11,000 meters followed by the lower stratosphere that ranges from 11,000 meters to 25,000 meters. Above 25,000 meters, the upper stratosphere model is used. Table \ref{ecrs_stand_atm} shows standard the parameterization of temperature and pressure as a function of vertical altitudes.
\begin{table}
  \centering
    \caption{Temperature and pressure variations for three different atmospheric regions: troposphere ($0 \! < \! h \! < \! 11,000$ m), lower stratosphere ($11,000 < h < 25,000$ m), and upper stratosphere ($h > 25,000$ m), are obtained from the NASA's Earth Atmospheric Model \cite{NASA_GRC}.}
    \begin{tabular}{| p{3.5cm} | p{4cm} | p{4cm} |} \hline
    \bf{Altitude, $h$ (m)}  & \bf{Temperature, $T$ ($^\circ$C)}  & \bf{Pressure, $p$ (kPa)}                           \\ \hline
     0 $<$ $h$ $<$ 11,000       &  15.04 - 6.49$\times$10$^{-3} h$   & $101.29 \times {[}\frac {T + 273.1}{288.08}{]}^{5.256}$   \\  \hline
    11,000 $<$ $h$ $<$ 25,000  &  -56.46                              & 22.65  e$^{1.73 - 0.000157 h}$                       \\ \hline
   $h$ $>$ 25,000             &  -131.21 + 2.99$\times$10$^{-3} h$  & $ 2.488 \times {[}\frac{T + 273.1}{216.6}{]}^{-11.388}$   \\ \hline
    \end{tabular}
    \label{ecrs_stand_atm}
\end{table}
Given the parameters described in Table~\ref{ecrs_stand_atm}, the air density, $\rho$ can be calculated using
\begin{equation}
	\rho = \frac{p} {0.2869 \times (T + 273.1)} ,
\label{ecrs_stand_atm_dens}
\end{equation}
where $\rho$ is in kg/m$^{3}$, temperature $T$ in $^\circ$C, and pressure $p$ in kPa.


One of the important features of the ECRS simulation is that the Earth is modeled at real-scale ($r_e = 6,371,200$ m) as a shell of water of 11,000 m in depth, which can be modified to accommodate other materials for specific applications. This also allows studying cosmic ray shower development beneath the earth surface.

\begin{figure}[htb]
\centering
\includegraphics[width=0.4\textwidth]{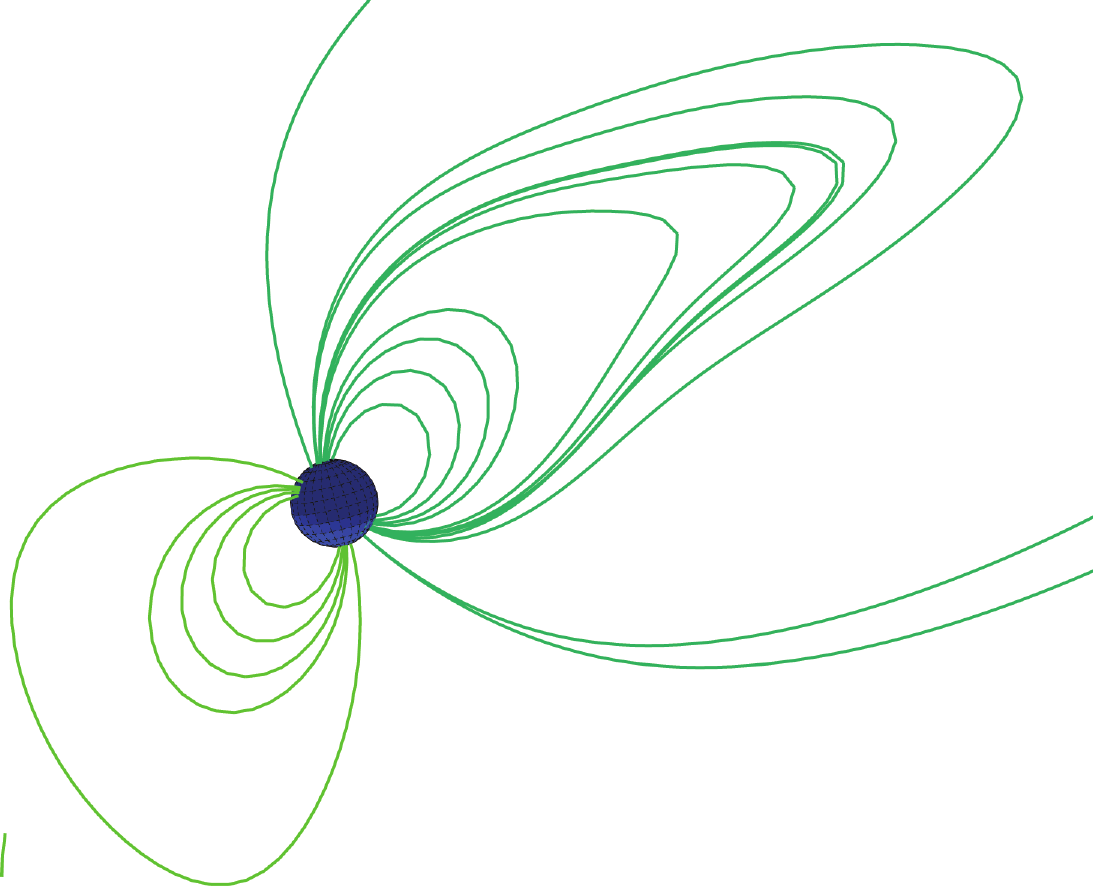} 
\caption{Visualization of the internal and external magnetic field implemented in ECRS.}
\label{magneticFieldLines}
\end{figure}

The geomagnetic field implemented in ECRS consists of the internal and external fields. The internal geomagnetic field is given by the IGRF model \cite{finlay:igrf}, and the external field is described by well established Tsyganenko models \cite{tsyganenko:model}.  
The Earth's external magnetic field is asymmetric with the Sun facing side being compressed due to solar wind and extends about 10 Earth's radii out while the other side stretches out in a magneto-tail that extends beyond 200 Earth's radii \cite{parks:space}, as shown in Fig.~\ref{magneticFieldLines}.

The main components of the primary cosmic rays that have galactic origin include energetic protons ($\sim$$74$\%) with the remaining being mostly alpha particles ($\sim$$70$\%) \cite{pdg}. The solar component of the primary cosmic rays consists predominately of protons, with a minor contribution from helium ions (about $10$\%). Most of these protons are of low energy ($\sim$$50$ MeV) and a smaller portion of solar proton that could have energies close to 1 GeV in association with distinct solar disturbances. These solar activities also modulate the low-energy galactic primaries that reach to the top of the Earth's atmosphere (11-year solar cycle).

\begin{figure*}
\centering
\includegraphics[width=0.85\textwidth]{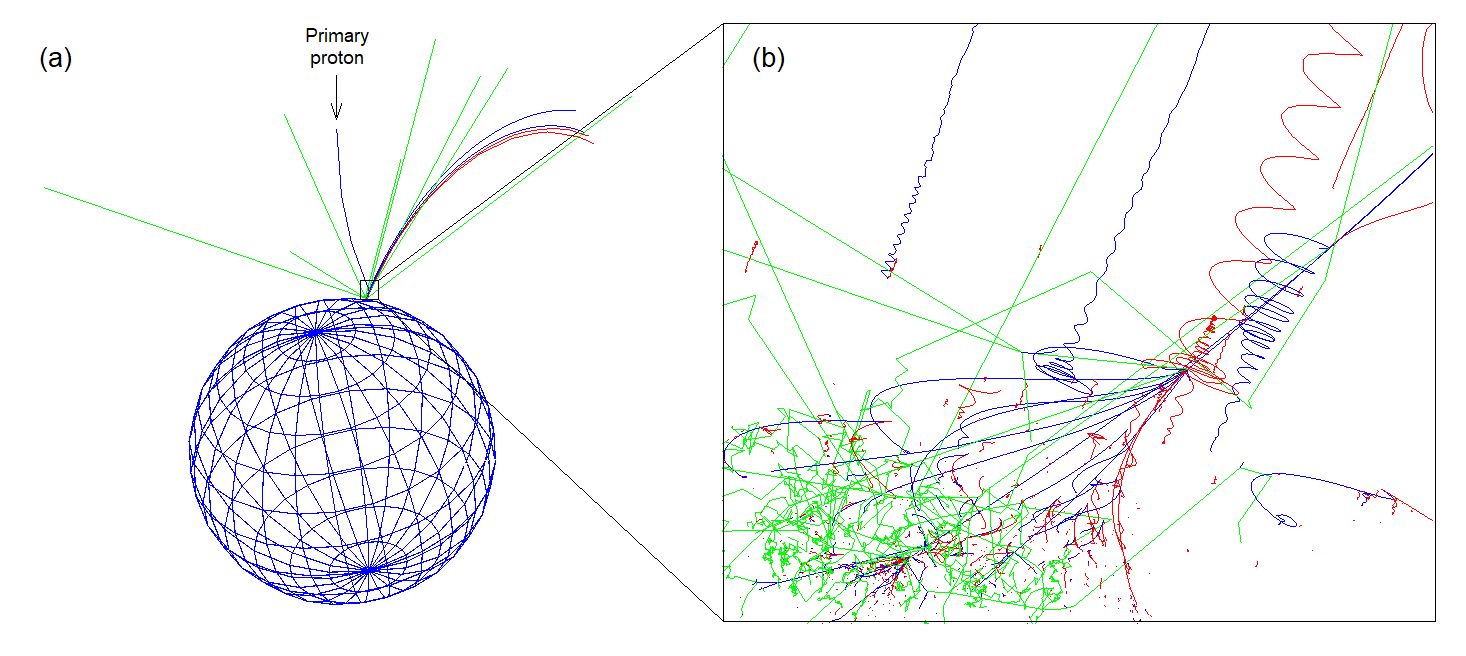}
\caption{A cosmic ray shower event display from a 100-GeV proton launched at a $1.2 R_e$ toward Atlanta area.
           (a) Distant view of the shower event. [Blue lines are the positive charged tracks; red lines the negative tracks and green lines the neutral particle (mostly gamma's)]
           (b) Zoomed-in view of the shower event which shows the curved trajectories from the charged tracks because of presence of the magnetic field. 
}
\label{magneticField}
\end{figure*}

In running the ECRS simulation, users can customize the primary particle type, energy, launch position, and direction. All secondary particles generated in the shower are recorded to an output file for later analysis, providing details such as their geo-coordinates, particle type, and momenta for subsequent offline data analysis. 
Figure \ref{magneticField} shows an example of cosmic ray shower event display from a $100$ GeV proton launched at a $1.2 R_e$ toward Atlanta, USA where the effect of the geomagnetic field is clearly demonstrated in the spiral trajectories. 

\begin{figure}
\includegraphics[width=0.95\textwidth]{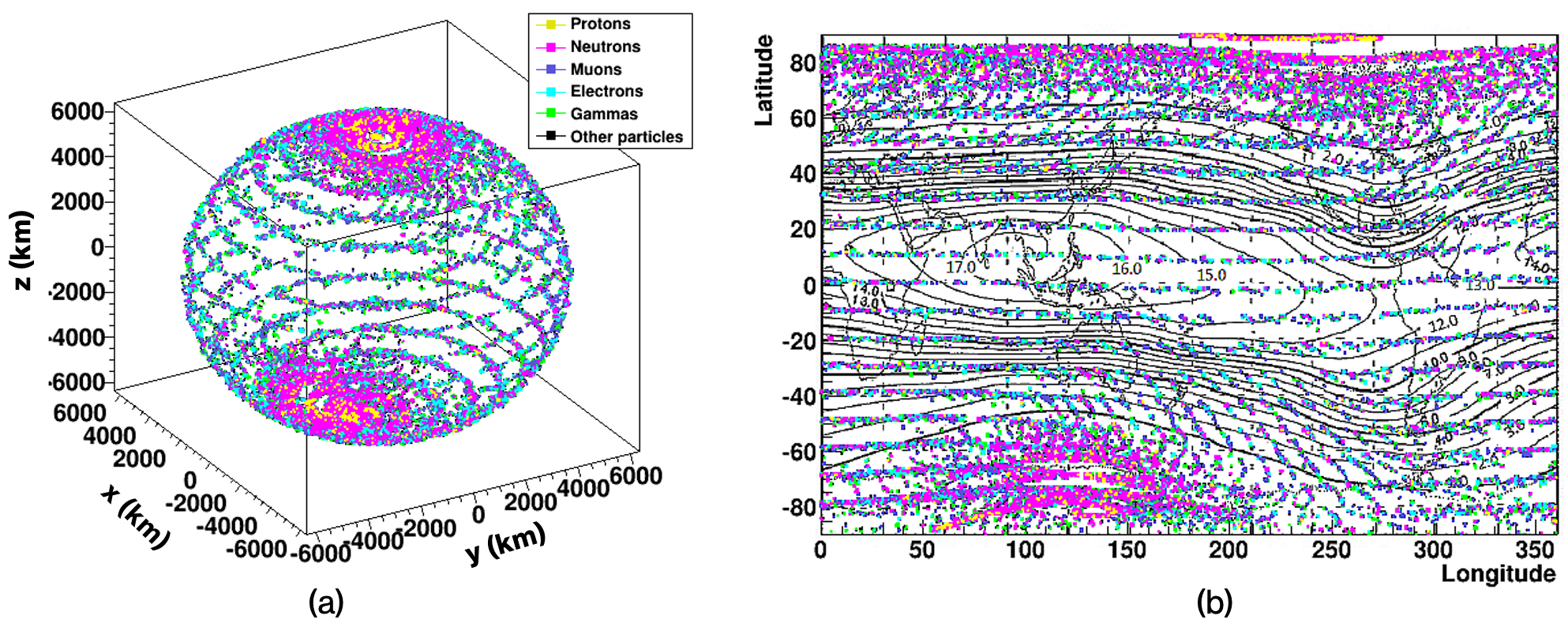}
\caption{ (a) A 3D display of the global particle distribution at the surface with magnetic field configuration of January 1, 2000. (b) Global map of geographical latitude versus longitude for the particles at the surface overplayed with the world grid of the vertical geomagnetic cutoff. The protons are in yellow, neutrons in magenta, muons in blue, electrons  in cyan, gammas in green, and all other particles are in black.}
\label{global_B}
\end{figure}

The primary cosmic rays are mainly comprised of charged particles that get deflected by the magnetic field when approaching the Earth. Without magnetic field effect, the particles at the surface will be evenly distributed over the global geographical coordinates. Because of the Earth's magnetic field, particles' distribution at the sea level highly depends on geoposition.
Figure~\ref{global_B}(a) shows a 3D global particle distribution at the surface by launching primary cosmic rays from $1.2 R_e$ toward the surface of the Earth at 10 degree increment in latitude and longitude. As one would expect, a greater concentration of the surface hits occurs at the poles where the cutoff rigidity is nearly zero while less cosmic ray particles reach to the equatorial region. 
In Figure~\ref{global_B}(b), the impact of the magnetic field on the distribution of particles at the surface overlaying the rigidity cutoff map is depicted. The geomagnetic cutoff rigidity serves as a numerical indicator of the protection offered by Earth's geomagnetic field. 

One of the advantages of the ECRS simulation is its ability to examine cosmic ray shower distributions globally following a specific geomagnetic configuration. This includes assessments at both atmospheric depths and sea level simultaneously. The outcomes of these simulations hold relevance for various interesting studies in the realm of cosmic ray applications as mentioned in Section \ref{sec:intro}. In the next section, we briefly summarize the results of two case studies of running the ECRS simulation.

\section{Case Studies} \label{sec:result}
The ECRS simulation offers a systematic approach to investigating the characteristics of global distributions of secondary cosmic shower particles. Its adaptability in configuring atmospheric air density and the geomagnetic field establishes it as a valuable tool for assessing variations in particle distributions across the entire atmospheric range, Earth's surface, and underground. For illustrative purposes, the following two subsections present case studies focusing on particle energy distributions at the Earth's surface.
\subsection{Global Distribution of Particle Energy at the Earth's Surface}
\label{global}
To quantify the characteristics of the cosmic ray shower particle distribution at global scale, we conducted the ECRS simulation on a computing farm (well over 10,000 compute cores) at the RHIC Computing Facility at Brookhaven National Laboratory. This involved launching 10,000 primary cosmic rays from 1.2 Earth's radii towards the Earth's surface with a 10 degree increment in latitude and longitude.  

The primary cosmic ray particles that are of interest for this study are protons with energies below 100 GeV which are dominant in the primary cosmic ray spectrum \cite{pdg}.   
The magnetic field was set to January 1, 2010 configuration which is close to the minimum of solar cycle 24.

%

\begin{figure*}[h]
\centering
\includegraphics[width=0.9\textwidth]{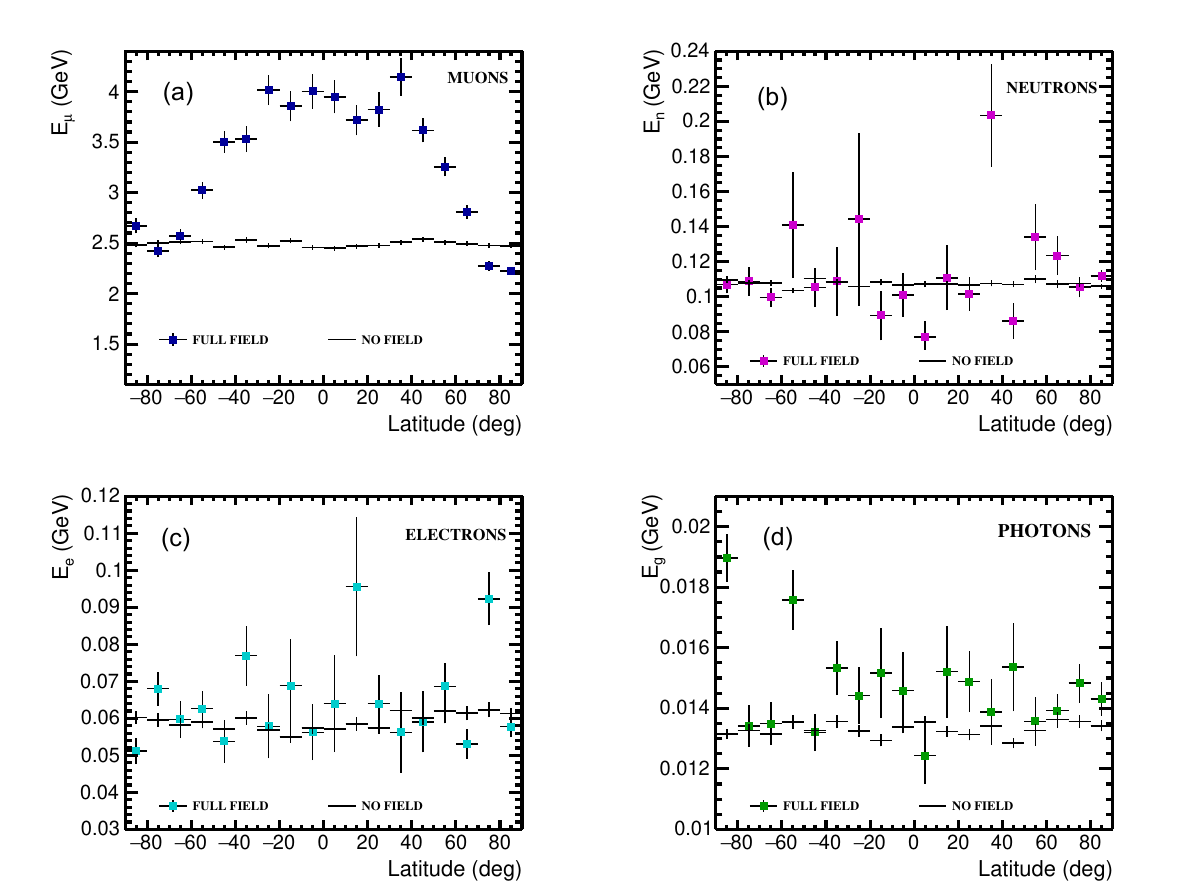}
  \caption{Latitudinal distribution of the particle energy at sea level. The vertical axis represents the mean particle energy as a function of geographic latitude integrated over all geographic longitudes. Primary cosmic rays were launched in 10 degree intervals in latitude and longitude with full geomagnetic field configured in January, 2010. Primary cosmic ray protons are launched from 1.2 Earth's radii with a purpose of studying the effect of the Earth's magnetic field.
}
\label{lat_energy}
\end{figure*}

Figure~\ref{lat_energy} shows the latitudinal distribution of the particle energy at sea level integrated over longitude direction. If there is no geomagnetic field, one would expect a flat energy distribution as represented by the dashed lines in Fig.~\ref{lat_energy}.
Panel (a) in Fig.~\ref{lat_energy} shows the average muon energy at sea level as a function of latitude. The average muon energy closer to the equator region is about 4 GeV within $\pm 20$ degrees in latitude and is steadily decreasing to 2.2 GeV toward the polar region as it is expected.
The variation of the neutron energy (in panel (b) of Fig.~\ref{lat_energy}) at the sea level fluctuates between 80 MeV and 140 MeV, which shows no strong variation in latitude. The data point at latitude about 35 degree is likely resulted from a large stochastic fluctuations in the simulated events. 
The electron energy at the surface uniformly varies from about 40 MeV to 100 MeV with little change in latitude similar to the neutron case, as shown in the panel (c) of Fig.~\ref{lat_energy}.
Figure~\ref{lat_energy}(d) shows the sea level energy of gamma ray photons as a function of latitude. The gamma ray energy varies from about 12 MeV to 16 MeV going from South to North Pole which is fairly constant on a global scale except two data points at latitude of about $-85$ and $-55$ degrees, respectively, with larger mean energy. These two data points are likely resulted from a large stochastic fluctuations in the simulated events as the case for neutron distribution. More computing resources will be needed to run the simulation with high statistics to confirm these observations.

\begin{figure*}[th]
\centering
\includegraphics[width=0.9\textwidth]{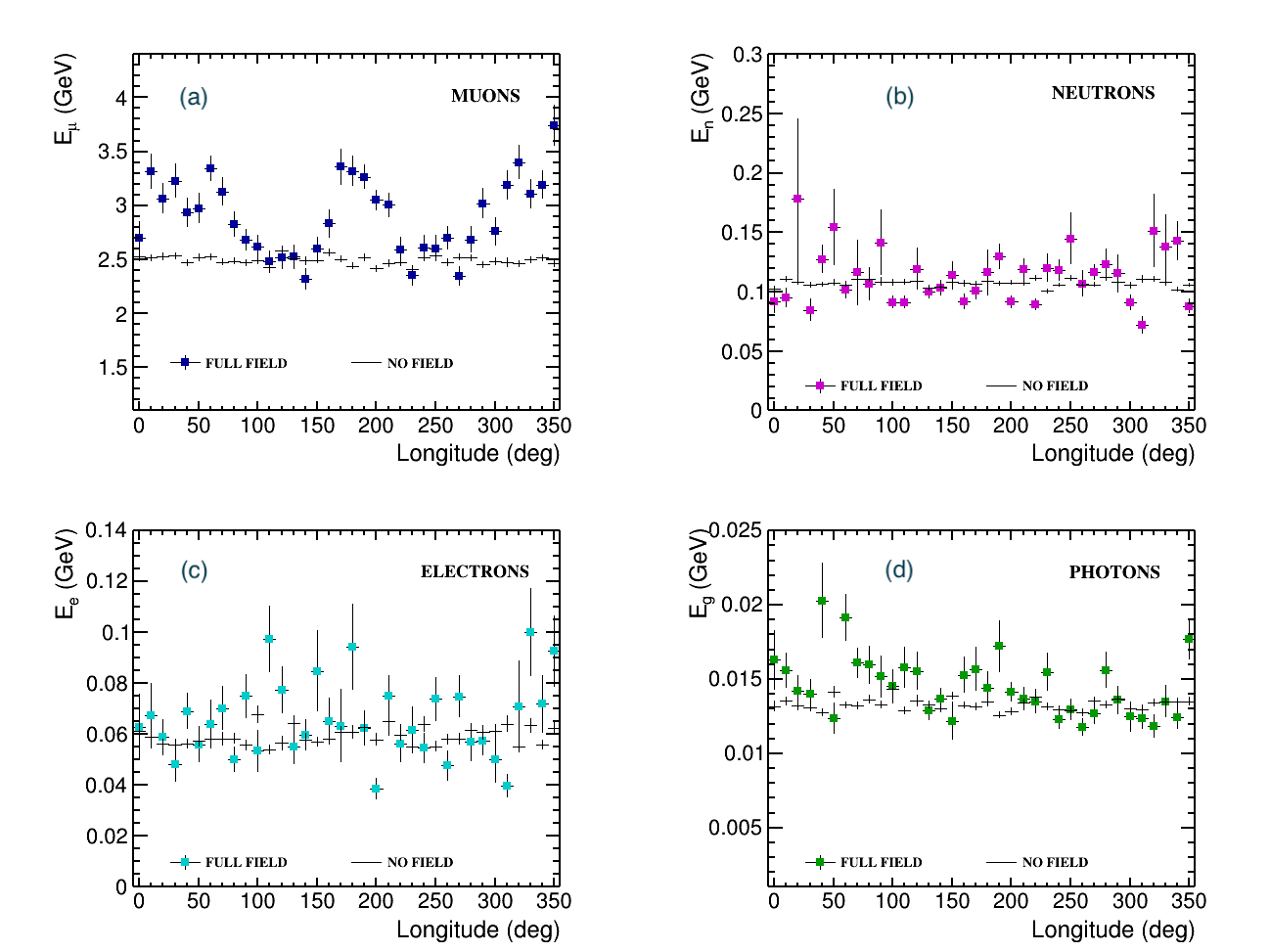}
\caption{Longitudinal distribution of the particle energy at sea level integrated over the latitude direction.
The dashed lines in each panel represents the mean value of the particle energy without magnetic field, in which case it is fairly constant over longitude. Primary cosmic rays were launched in 10 degree intervals in latitude and longitude using January, 2010 geomagnetic field configuration. Primary cosmic ray protons are launched from 1.2 Earth's radii with a purpose of studying the effect of the Earth's magnetic field.
}
  \label{long_energy}
\end{figure*}

We also look at the longitudinal variations of the secondary cosmic ray particles at sea level integrated over the latitude direction. The results are shown in Fig.~\ref{long_energy}.
The average of muon energy as a function of geographic longitude shows a sinusoidal pattern as traveling around the globe with energy varying between $\sim$4 GeV and 2 GeV as shown in Fig.~\ref{long_energy}(a), which is associated with the asymmetry of the geomagnetic field as it is shown in Fig.~\ref{magneticFieldLines}.
Figure~\ref{long_energy}(b) shows the neutron energy distribution as a function of longitude. The energy of neutrons is fairly steady around the globe, varying between about 80 MeV to 140 MeV.
The electron energy variation at the surface with longitude is shown in Fig.~\ref{long_energy}(c). There is a small energy variation from as low as $\sim$40 MeV to 100 MeV without apparent pattern in energy as the longitude changes.
The energy variation of gamma ray photons at the surface is minimal with longitude, changing between about 10 MeV to 20 MeV as shown in Fig.~\ref{long_energy}(d).

The results shown in Fig.~\ref{lat_energy} and Fig.~\ref{long_energy} demonstrate the characteristic average energies of muons, neutrons, electrons, and photons from cosmic ray showers at sea level for a given geomagnetic field configuration (Januray 1, 2010). In contrast to neutrons, electrons and photons, the mean energy of muons exhibits a more pronounced dependence on both geographic latitude and longitude. This intriguing pattern observed in the average muon energy (as well as the corresponding counts) introduces a novel avenue for monitoring dynamic changes in the geomagnetic field on a global scale.

\subsection{Cosmic Ray Muon Energy Variations at Selected Locations}
\begin{figure*}[th]
    \centering
    \includegraphics[width=0.80\textwidth]{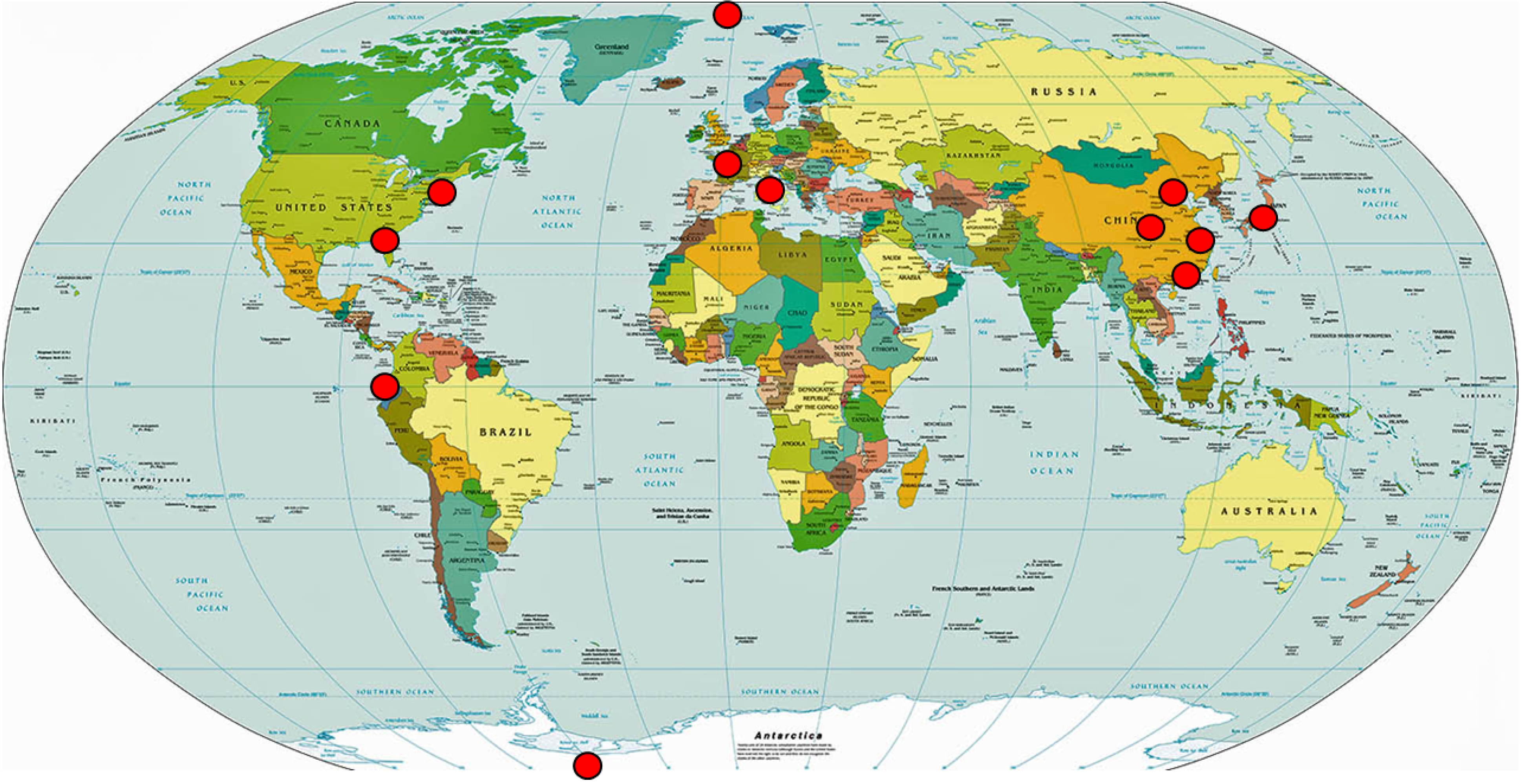}
    \caption{Selected geopolocations (marked in red) for performing ECRS simulation. The figure is from the author's dissertation~\cite{OlesyaSarajlic2017}.}
    \label{fig:map}
\end{figure*}

One of the flexible features of running ECRS is the ability to customize the simulation with a specified geomagnetic configuration. This allows for the isolation of the direction of incoming primary cosmic rays, facilitating the study of characteristics in the distribution of secondary cosmic ray particles at chosen geopositions. As a demonstration, ECRS simulation was carried out for several major cities around the world, which include
New York, USA (40.71$^\circ$ N, 285.99$^\circ$ E, R$_{c}$ = 2.4 GV); 
Atlanta, USA (33.75$^\circ$ N, 275.61$^\circ$ E, R$_{c}$ = 3.6 GV); 
Paris, France (48.86$^\circ$ N, 2.35$^\circ$ E, R$_{c}$ = 3.7 GV); 
Rome, Italy (41.90$^\circ$ N, 12.50$^\circ$ E, R$_{c}$ = 6.3 GV); 
Beijing, China (39.90$^\circ$ N, 116.41$^\circ$ E, R$_{c}$ = 8.8 GV); 
Nagoya, Japan (35.18$^\circ$ N, 136.91$^\circ$ E, R$_{c}$ = 11.2 GV); 
Ecuador, South America (-1.83$^\circ$ N, 281.82$^\circ$ E, R$_{c}$ = 12.2 GV); 
Shanghai, China (31.23$^\circ$ N, 121.47$^\circ$ E, R$_{c}$ = 13.2 GV); 
Xi'an, China (34.34$^\circ$ N, 108.94$^\circ$ E, R$_{c}$ = 14.1 GV); and 
Hong Kong, China (22.40$^\circ$ N, 114.11$^\circ$ E, R$_{c}$ = 15.6 GV).
For a comparison, ECRS was also run for North Pole (90.00$^\circ$ N, R$_{c}$ = 0.1 GV) and 
South Pole (-90.00$^\circ$ N, R$_{c}$ = 0.1 GV) locations. These sites are marked in red in the world map as shown in Fig.~\ref{fig:map}.

\begin{figure}[th]
\centering
\includegraphics[width=0.48\textwidth]{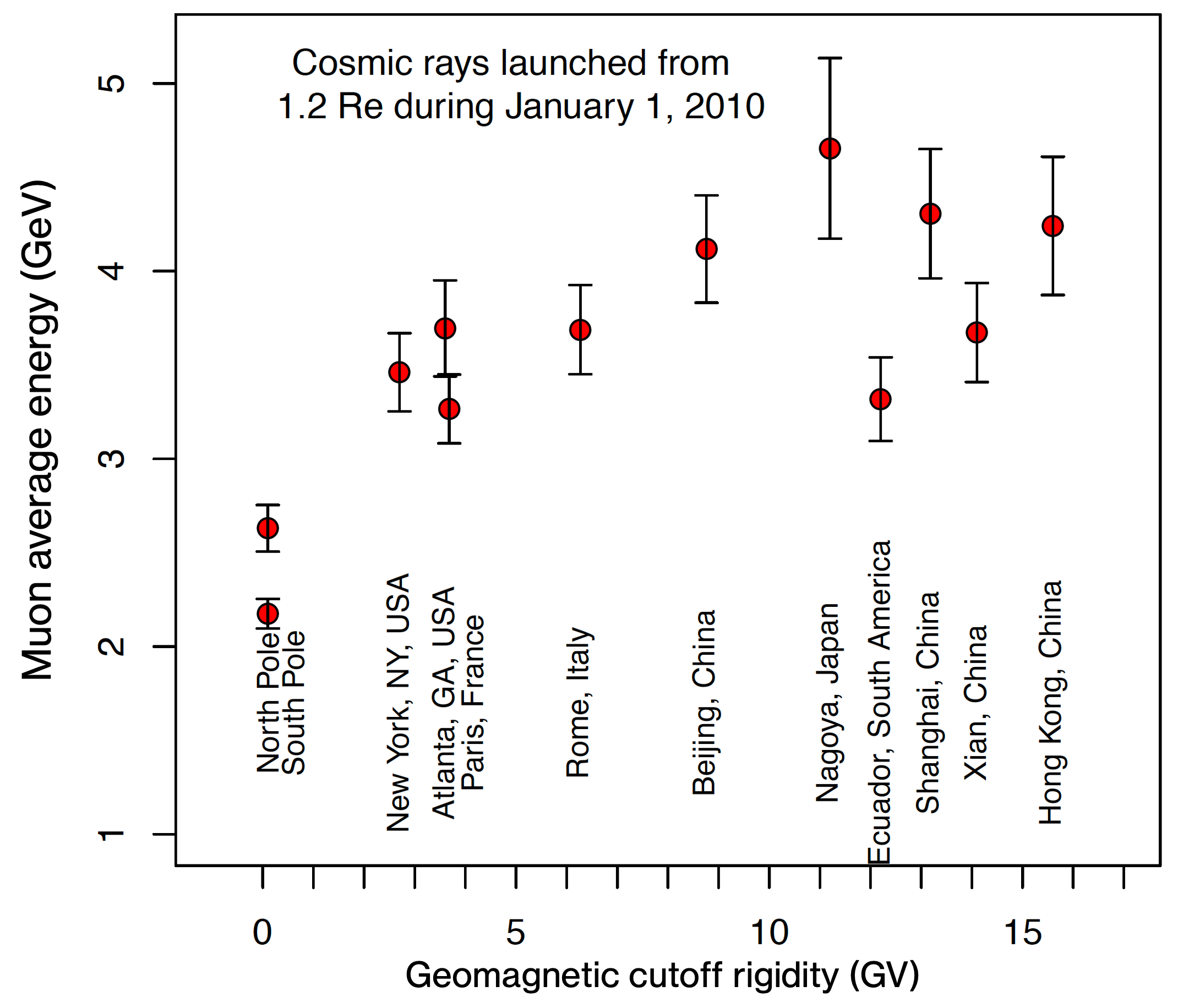} 
\caption{ The average muon energy at selected cities as a function of the geomagnetic cutoff rigidity with the magnetic field configuration during January of 2010. Primary cosmic ray protons are launched from 1.2 Earth's radii~\cite{OlesyaSarajlic2017}. 
}
\label{CitiesAll_KEmu_Rc_err_lab}
\end{figure}

Figure~\ref{CitiesAll_KEmu_Rc_err_lab} shows the average muon energy at the ground level as a function of geomagnetic cutoff rigidity in GV. As it is expected, the average muon energy increases from $\sim$2 GeV to 4 GeV when transitioning from the Polar region towards the equatorial plane with the cutoff rigidity.

As one would expect, running ECRS simulation can be compute intensive for tracking every secondary particles created in cosmic ray showers in the magnetic field in the entire atmospheric depth. A previous study on computing needs and challenges of running ECRS was reported in \cite{pearc18}. 
With the increase in compute availability and compute power on campus and though national projects like the XSEDE~\cite{XSEDE} (followed by ACCESS~\cite{ACCESS}) and Open Science Grid~\cite{OSG1,OSG2}
over the past several years, certain challenges, particularly the duration of running time, are no longer a significant concern to conducting high-statistics ECRS simulations in a high performance/throughput computing (HP/TC) cluster environment.

\section{Summary and Outlook} \label{sec:summary}

With the increasing global interest in utilizing cosmic ray measurements for practical applications, spanning cosmic ray particle tomography, space and terrestrial weather monitoring, area-averaged soil moisture monitoring, to addressing public health concerns related to cosmic ray radiation exposure during flights, there is a demand for a simulation tool. This tool should feature flexible, configurable parameters to systematically characterize the properties of cosmic ray showers on a global scale.  In this study, we presented such a tool with detailed 
description of cosmic ray shower simulation based on GEANT4. This simulation enables a more accurate estimation of cosmic ray radiation throughout the Earth's atmosphere, considering the complex interactions of cosmic rays with the atmosphere, magnetic field, and other environmental factors.

As it is presented in Section~\ref{global}, cosmic ray muon energy and count at the ground level vary with geoposition at a given magnetic field. 
The simulated sensitivity of muon variation to changes in the magnetic field can be integrated with the measured muon flux variation from a global network of detectors. This combination allows for the development of a novel ground-based system to monitor dynamic changes in space weather~\cite{https://doi.org/10.1029/2023JA031943}.

Another potential use of ECRS is to calculate the muon weighting function in the region of upper troposphere/lower stratosphere at different geographic locations for studying the effective atmospheric temperature at large scale using ground-level muon measurements~\cite{XiaohangZhang:2016}.

Looking forward, 
as computing resources continue to advance, running ECRS simulations will become more practical for numerous interesting cosmic ray applications. There is an ongoing plan to release the ECRS code on GitHub for general use, aligning with the release of the new version of GEANT4.



\acknowledgments
The authors would like to acknowledge the support of the study under the RISE program at Georgia State University.

\bibliographystyle{JHEP}
\bibliography{main}



\end{document}